\documentclass[12pt]{article}
\usepackage[latin1]{inputenc}
\usepackage{pstricks,pst-node,pst-text,pst-3d}
\usepackage{graphicx}
\begin{document}
\begin{center}
{\Huge Some surprising results of the Kohn-Sham Density Functional}
\end{center}
\vspace{2cm}

L. G. Ferreira$^1$, M. Marques$^2$, L. K. Teles$^2$, R. R. Pel\'a$^2$

$^1$Instituto de F\'{\i}sica, Universidade de S\~ao Paulo, Brazil

$^2$Instituto Tecnol\'ogico de Aeron\'autica, S\~ao Jos\'e dos Campos, Brazil
\section{abstract}
For some insulators we present a procedure to determine an electronic density leading to a lower energy than that of the
Kohn-Sham ground state.
\section{Introduction}
The LDA-1/2 method, for the calculation of semiconductor band gaps, is well known nowadays \cite{nois,nois2}. It is simple, fast and precise, with a precision competing with that of GW \cite{Hedin}, though orders of magnitude faster 
\cite{Mauro,OP,Furth,matu}.  The method has a very long story because it is based on the half-occupation technique
for the calculation of atomic ionization potentials \cite{Slater,Leite,Janak}.
\par The standard KS many-electron Hamiltonian is \cite{Kohn}
\begin{equation}
E[n]=K[n]+U_{pn}[n]+U_{nn}[n]+X_c[n]
\end{equation}
where $X_c$ is the exchange-correlation term, not necessarily LDA \cite{LDAca,LDAca2}. The LDA-1/2 Hamiltonian is written as
\begin{equation}
\Xi[n]=E[n]+\int V_Sn
\end{equation}
where $V_S$, typical of the method, is named {\it self-energy potential} and is obtained from simple atomic calculations.
Clearly
\begin{equation}
\left . \frac{\delta \Xi}{\delta{V_S}}\right|_{V_S=0}=n
\label{derivative}
\end{equation}
On minimizing $\Xi[n]$ with respect to the wavefunctions we obtain the KS equations but with the added external field
$V_S$. In the self-consistent solution, the wave functions define the electron number-density $n$ which gives a total energy
$\Xi[n]$ in no way equal to the ground state energy $E[n]$. These authors never failed to insist that the LDA-1/2 method could give correct one-electron energies (because it includes the self-energy) but could not be expected to give total energies, lattice parameters, etc.

In this work we decided to investigate the total energy calculated as
\begin{equation}
E[n]=\Xi[n]-\int V_Sn
\end{equation}
A first version of this work was presented in the 2013 March Meeting \newline
(http://meetings.aps.org/link/BAPS.2013.MAR.T24.7). At the time we presented the results calculated with code VASP\cite{VASP}. The present results are calculated with the all-electron WIEN2k-LAPW\cite{Blaha}. We found our results so surprising that we delayed their publication. 

Before presenting results we remind what should be expected. Let $n_0$ be the number-density obtained by minimizing $E[n]$ and let $\eta$ be the number-density obtained by minimizing $\Xi[n]$. They are different because of the
self-energy potential $V_S$. Of course one expects
\begin{equation}
E[n_0]<E[\eta]
\end{equation}
because $n_0$ is the density minimizing $E$. Surprise: this does not happen for most III-V semiconductors. That raises an important question: what is the meaning of $\eta$, and why it gives an energy lower than the minimum.

In order to explain our figures we must remind an important feature of LDA-1/2. The self-energy potential is
\begin{equation}
V_S=CUT\_function\times[V(1/2\  ion)-V(atom)]
\end{equation}
where $V(atom)$ and $V(1/2\  ion)$  are the atomic and the ionic potential with 1/2 electron removed. The $CUT\_function$ has to be introduced because the difference between the two potentials has a Coulomb tail $1/r$ at infinity. If this potential is to be repeated in a infinite lattice the calculation will diverge. So in what will follow we will present figures of $E[\eta]$ and $gap$ as functions of the parameter $CUT$ of the function.

\section{Some DF05 results}
\begin{figure}[h]
\begin{center}
\includegraphics[scale=0.5]{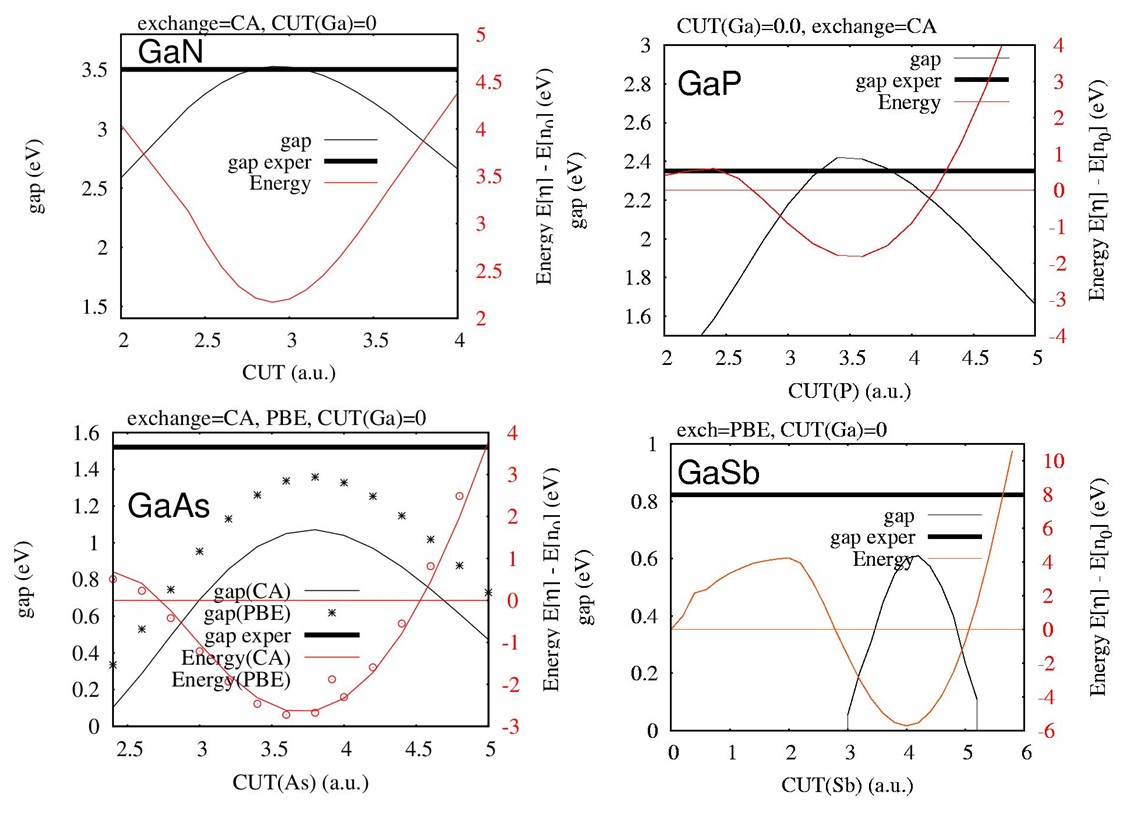}
\caption{Some III-V semiconductors where one observes regions of $E[n_0]>E[\eta]$, in apparent violation of
the Density Function theory. In the case of GaAs we present results for both PBE and
LDA (CA) exchange-correlations. We use both exchange correlations because they do not lead to qualitative differences.
In the case of GaN there is no negative region but only a minimum.}
\label{examples}
\end{center}
\end{figure}

\begin{figure}[h]
\begin{center}
\includegraphics[scale=0.5]{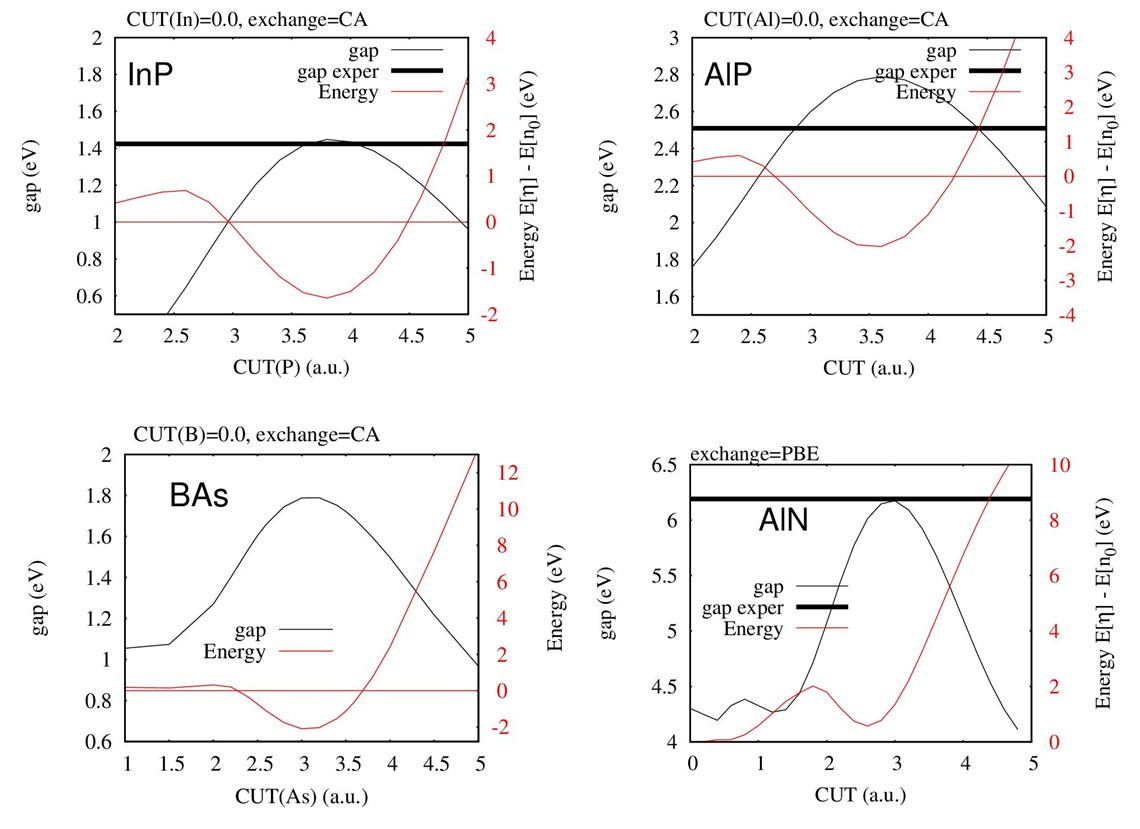}
\caption{ Other III-V's for which one observes the unusual behaviour. For the nitride, again there is no negative region but only a minimum.}
\label{examples2}
\end{center}
\end{figure}

\begin{figure}[h]
\begin{center}
\includegraphics[scale=0.45]{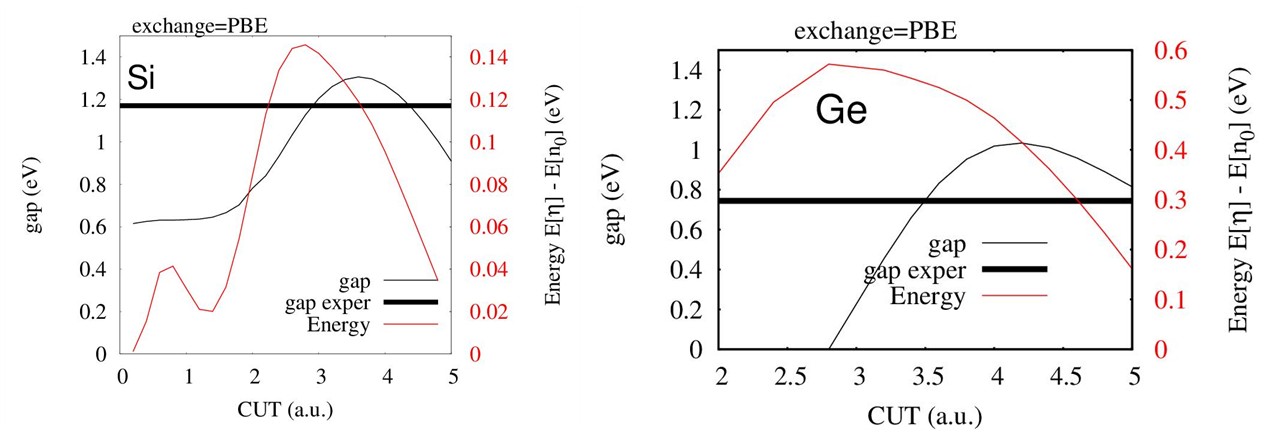}
\caption{ In the case of Silicon and Germanium there is no region where  $E[n_0]>E[\eta]$. We like to think that this is associated with the indirect gap of these materials.}
\label{examples3}
\end{center}
\end{figure}

First we have to name our method to calculate total energies. We abandon the name LDA because the calculation  might be GGA 
\cite{GGA} as well. So we use ``DF'' and maintain the 05 to remind us of the origin, which is the removal of 1/2 electron 
\cite{Slater,Leite,Janak}. In no way the 05 is a version of the method. In Figs.~\ref{examples}, ~\ref{examples2} and \ref{examples3}  we
present some examples of how gap and the energy $E[n]=\Xi[n]-\int V_Sn$ depend on the parameter $CUT$ of the $CUT\_function$. The examples are chosen among some III-V semiconductors where we observe a region of $CUT$ where $E[n_0]>E[\eta]$, and Si and Ge, where this region is not observed. A common feature of the III-V's and other semiconductors presenting the same behaviour is that the minimum of the total energy happens at the same value of $CUT$ of the maximum gap. The gaps presented in the  Figs. are not always equal to those of the LDA-1/2 papers because here we are not adding a self-energy potential for the cation.
\par The results presented in the Figure are surprising, and at a first glance seem to violate the DF theorems. This point is discussed in a latter section. Further, the simultaneous appearance of a maximum gap and a total energy minimum should be no surprise. In section \ref{Penney} we show this also happens in the simplest periodic model, the Kronig-Penney model.
A remarkable fact about the minima of $E[\eta]$ is that if we start self-consistent interactions from $\eta$ we do not reach a new stationnary point but return to $n_0$. In the last section of this paper we present  a possible explanation for this behaviour.
\section{The Kronig-Penney model}\label{Penney}
\begin{figure}[h]
\begin{center}
\includegraphics[scale=0.6]{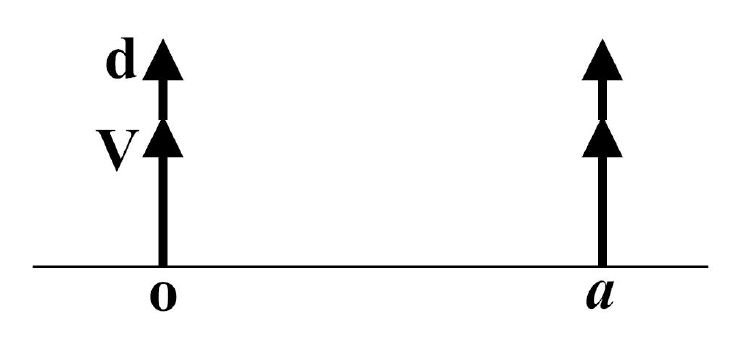}
\caption{Kronig-Penney model with delta-function potentials. $d$ is a perrurbation, analogous
to the self-energy potential $V_S$ of 3D crystals.}
\label{Kronig}
\end{center}
\end{figure}

\begin{figure}[h]
\begin{center}
\includegraphics[scale=0.5]{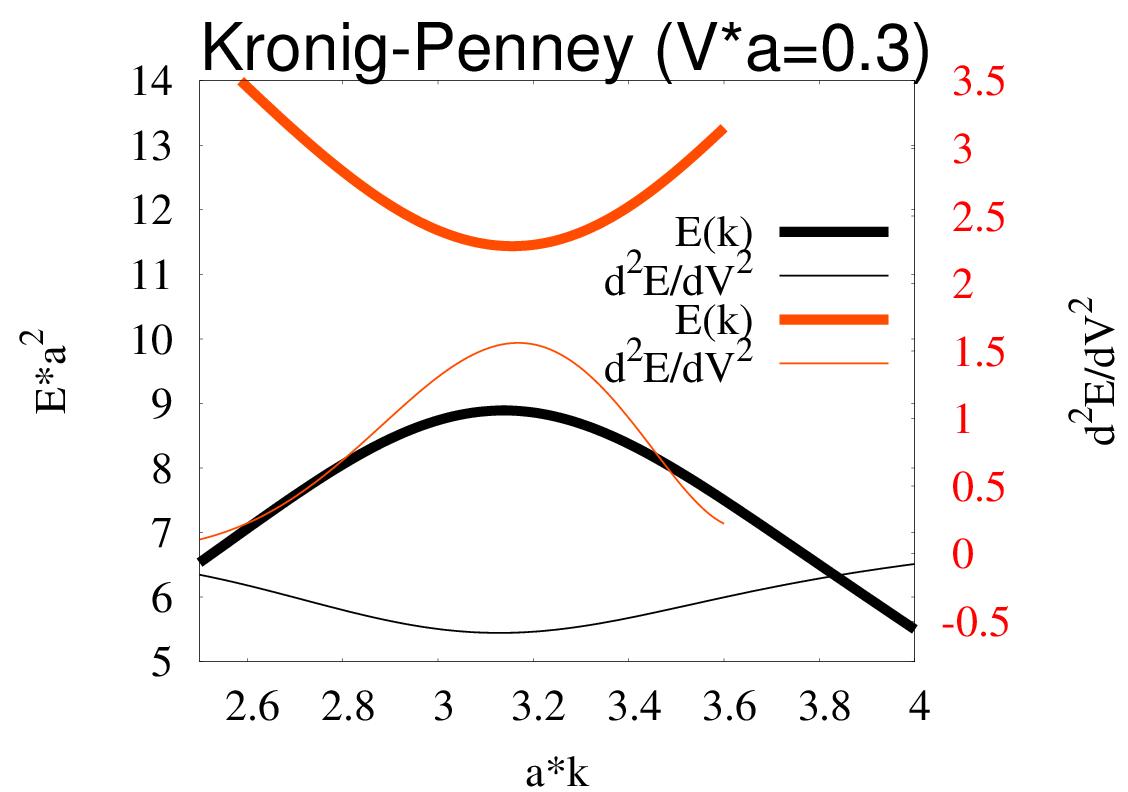}
\caption{The black lines refer to the valence band and the red to the conduction band. The thick lines are band functions and the thin lines are second derivatives, positive for the conduction and negative for the valence.}
\label{Kronig2}
\end{center}
\end{figure}

The best way to describe the simultaneous opening of the band gap and the lowering of the energy is by recurring to the Kronig-Penney model. Consider the model one-dimensional potential of Fig.\ref{Kronig} made of periodic delta-functions of amplitude $V$ spaced with lattice parameter $a$.
$d$ is a perturbation analogous to the self-energy potential $V_S$ of 3D crystals. The subtraction of $\int V_Sn$ is equivalent to removing the first-order perturbation of $V_S$ (see Eq.~\ref{derivative}). There remains second and higher-orders perturbations. Accordingly, in the case of the Kronig-Penney model we calculate the second-order perturbation $d^2E/dV^2$ in the vicinity of the band gap at $k=\pi/a$. The result is represented in Fig.\ref{Kronig2}, where we see that the perturbation opens the gap and lowers the valence band energies.
What happens is simple to understand: the perturbation increases the gap by lowering the valence and raising the conduction.
In the 3D case, the $V_S$ perturbation might raise the condction and lower the valence, decreasing the total energy. The 3D case is a lot more complicate than the Kronig-Penney model. For instance, the perturbation on Si does not decrease its total energy and the band gap does not happen at the maximum of the valence band. Of course Si is an indirect gap semiconductor and the maximum of the valence band is not at the same point as the minimum of the conduction band.
\section{Violating DF theorems?}
\label{theorems}
A naive explanation for $E[n_0]>E[\eta]$ is to blame the exchange-correlation approximations being used. The case of 
GaAs presented in Fig.\ref{examples} suggests that the result will happen for any approximation. The fact that we found a non-stationnary 
density $\eta$ having a lower energy than $n_0$ can only mean that, in varying the density aiming at an energy minimum we 
restrited the variation severely. The restriction that is always made in solid calculations is to impose a certain translation 
symmetry by defining a priori the unit cell. If we found a lower energy non-stationnary state means that the true ground state 
has a larger unit cell or it is not a truly periodic system (like Chromium, for instance). In the case of the III-V's, the ground
 state might include a stationnary charge-density wave of transfer to and from the cation or anion, in the
same way as in magnetic systems where we may have stationnary spin densities.

\begin{figure}[h]
\begin{center}
\includegraphics[scale=0.5]{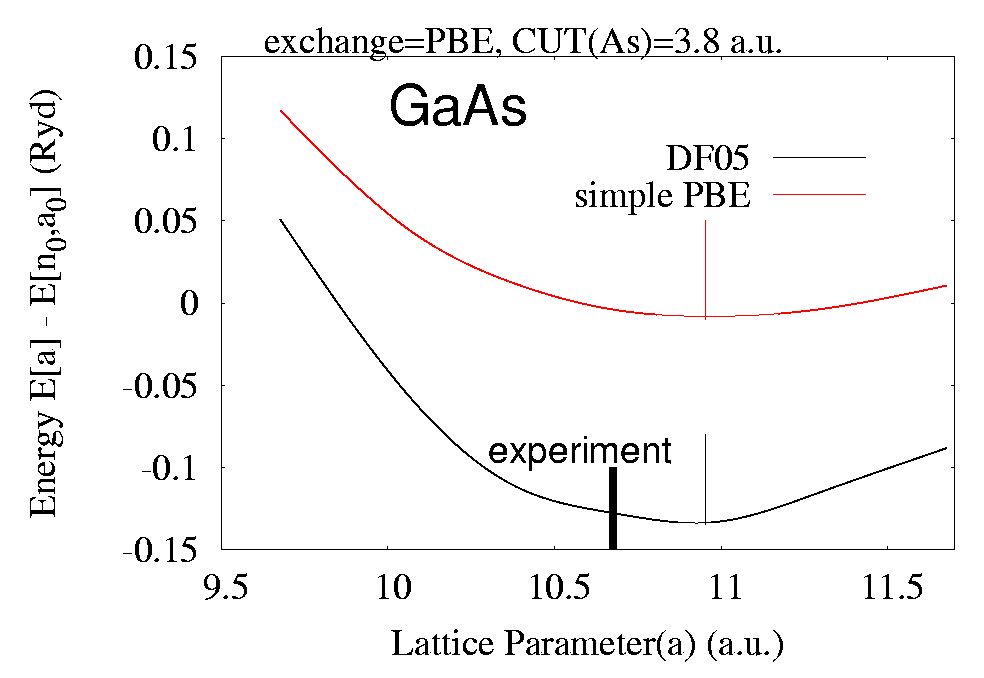}
\caption{GaAs. Energies as functions of the lattice parameter a. For each curve, the minimum energy is marked.}
\label{lattice}
\end{center}
\end{figure}

One might argue that the error is not in the {\it a priori} symmetry but in the assumed lattice parameter. All our calculations were made  at the experimental lattice parameter and they might be inadequate for the  energy calculations. To bar this
possibility we present in Fig.\ref{lattice} a study of energy versus lattice parameter, both for the standard GGA
exchange-correlation and for DF05 (also with GGA exchange-correlation). One notices that the energy minima happens at lattice sizes very close to experiment and the energy differences due to size are negligible compared the difference between DF05 and standard GGA.

\begin{figure}[h]
\begin{center}
\includegraphics[scale=0.5]{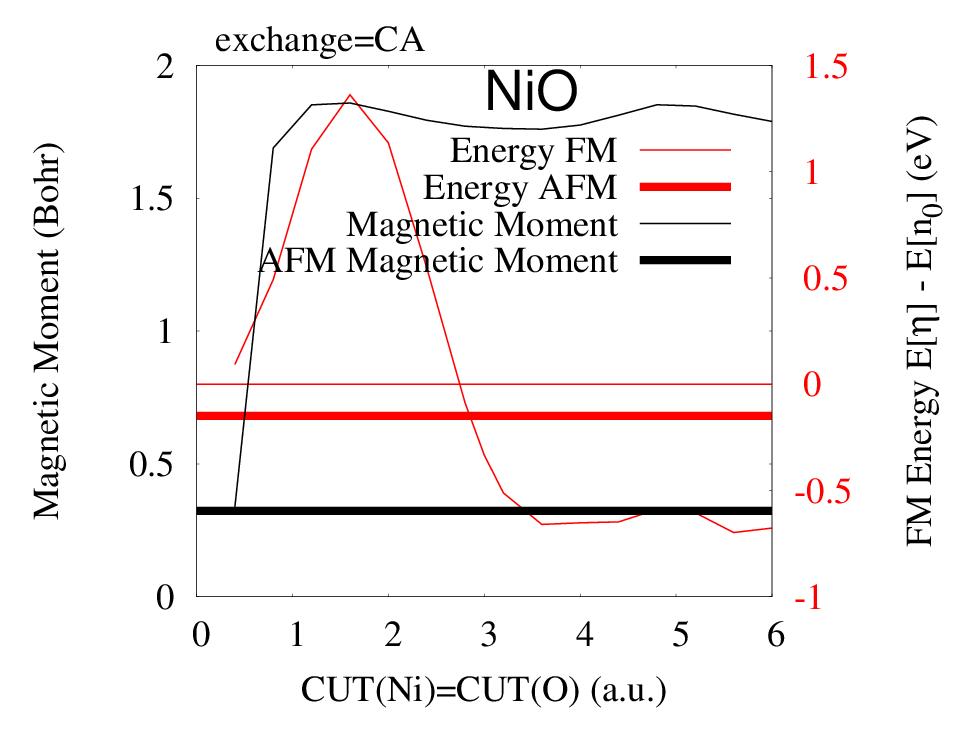}
\caption{NiO. Energy and Magnetic Moment of the standard AFM solution compared with the DF05 solution.}
\label{NiO}
\end{center}
\end{figure}

The translation symmetry breaking is well illustrated by the antiferromagnets. In this case the full symmetry of the lattice breaks
into at least two sublattices: one with spin up and the other with spin down. The case of the antiferromagnetic NiO is explained 
in Fig.\ref{NiO}. The standard solution of NiO assumes a rhombohedral lattice with two Ni per cell. The lattice is almost fcc, 
the distortion being due to the spin interactions. To apply the DF05 procedures to magnetic NiO we have to generalize the self-energy 
potentials to include spin-polarization. We begin with the standard solution for NiO, assuming a rhombohedral lattice with two Ni atoms 
(and 2 Oxygens) per cell. No self-energy potential $V_S$ is added. The self-consistent solution is indeed antiferromagnetic, but with a sublattice magnetic 
moment too small (only $0.3 Bohr$ against $\simeq 2$ from experiment). The energy of the AFM solution is below the zero of the standard 
FM solution, indicating that AFM is indeed the ground state. Now we add the spin-dependent self-energy potential to the FM arrangement (only one Ni per cell), and, after 
self-consistency, subtract $\int nV_S$. As function of the parameter $CUT$, the energy first increases then drops to a level below
 the AFM solution. The sublattice magnetization reaches a level very similar to experiment. Does this calculation means that NiO is 
 ferromagnetic? By no means, it only implies the existence of a solution, very likely antiferromagnetic, with much larger unit cell

\section{Conclusions}
This paper brings an apparent difficulty to DF theory in that we find all-electron states with lower energy than the official DF ground state. As we argue, the difficulty is only apparent because we usually do not explore the whole wavefunction space by restricting the translational symmetry. In this sense, some of our results are worth noticing, for instance the true ground state of the III-V's is not zinc-blende, which is universally assumed.

\end{document}